\documentclass[12pt,a4paper,twocolumn]{article}

\usepackage[british]{babel}

\usepackage[a4paper,top=2cm,bottom=2cm,left=2.5cm,right=2.5cm,marginparwidth=1.75cm]{geometry}

\usepackage[style=numeric,sorting=none]{biblatex}
\DeclareFieldFormat[article]{volume}{\textbf{#1}}

\usepackage{amsmath}
\usepackage{graphicx}
\usepackage[colorlinks=true, allcolors=blue]{hyperref}
\usepackage{hyperref}
\usepackage[title]{appendix}
\usepackage{mathrsfs}
\usepackage{amsfonts}
\usepackage{booktabs} 
\usepackage{threeparttable} 
\usepackage{algorithm}
\usepackage{algorithmicx}
\usepackage{algpseudocode}
\usepackage{listings}
\usepackage{enumitem}
\usepackage{chngcntr}
\usepackage{booktabs}
\usepackage{lipsum}
\usepackage{subcaption}
\usepackage{authblk}
\usepackage[T1]{fontenc}    
\usepackage{csquotes}       
\usepackage{diagbox}
\usepackage{comment}
\usepackage{url}
\usepackage{lmodern}

\usepackage[justification=raggedright, singlelinecheck=false]{caption}
\usepackage{helvet}  

\usepackage{setspace}
\usepackage{titlesec}
\titleformat{\section} 
  {\normalfont\Large\bfseries}{\thesection.}{1em}{}
  
\usepackage{float}   
\usepackage{caption}


\date{}  

\begin{document}

\twocolumn[

{\Large \bfseries A Software-Defined Radio Testbed for Distributed LiDAR Point Cloud Sharing with IEEE 802.11p in V2V Networks\par}
\vspace{2ex}

\begin{center}
    {\large Mario Hernandez\textsuperscript{1} \quad Elijah Bryce\textsuperscript{3} \quad Peter Stubberud\textsuperscript{2}} \\
    {\large Ebrahim Saberinia\textsuperscript{2} \quad Brendan Morris\textsuperscript{2}} \\
    \vspace{1ex}
    {\small \textsuperscript{1} University of Puerto Rico, RP \textsuperscript{2} University of Nevada, Las Vegas \textsuperscript{3} Las Positas College} \\
    {\small mario.hernandez4@upr.edu \quad elijahjbryce@gmail.com \quad peter.stubberud@unlv.edu} \\
    {\small ebrahim.saberinia@unlv.edu \quad brendan.morris@unlv.edu}
\end{center}

]

\begin{abstract}
We present a Software Defined Radio (SDR)–based IEEE 802.11p testbed for distributed Vehicle-to-Vehicle (V2V) communication\footnote{Code: \href{https://github.com/pollyjuice74/V2V-SDR-Networking-Testbed}{\texttt{here}}.}. The platform bridges the gap between network simulation and deployment by providing a modular codebase configured for cost-effective ADALM-Pluto SDRs. Any device capable of running a Docker with ROS, executing Matlab and interface with a Pluto via USB can act as a communication node. To demonstrate collaborative sensing, we share LiDAR point clouds between nodes and fuse them into a collective perception environment. We evaluated a theoretical model for leveraging decentralized storage systems (IPFS and Filecoin), analyzing constraints such as node storage convergence, latency, and scalability. In addition, we provide a channel quality study.
\end{abstract}

\section{Introduction}

Vehicle-to-Vehicle communication is a key enabler of intelligent transportation systems and autonomous driving. By communicating environmental information, vehicles can improve situational awareness, enhance safety and coordinate maneuvers. Standards such as IEEE 802.11p provide the physical (PHY) and medium access control (MAC) layers needed for Dedicated Short-Range Communication (DSRC), but significant challenges remain in evaluating real-world performance of perception sharing across heterogeneous nodes.

Recent advancements in blockchain-based storage such as InterPlanetary File System (IPFS) and Filecoin have introduced decentralized, incentivized, tamper-resistant methods for sharing large datasets across distributed networks. This addresses a critical safety issue in these multi-node infrastructure networks, where data could be easily read and tampered with by a knowledgeable party. 

For V2V communication systems, these decentralized peer-to-peer storage methods offer the potential to offload and archive high-bandwidth perception data, including LiDAR point clouds, while ensuring verifiability, resilience against single points of failure, and availability to nodes without direct internet connectivity.

Our main contributions are as follow:
\begin{itemize}[itemsep=1pt, topsep=2pt] 
    \item We developed a Dockerized, lightweight ROS base node that is fully extensible, enabling heterogeneous devices to join the network with minimal setup.
    \item We integrate MATLAB-based SDR communication with ROS, enabling real-time fusion of LiDAR point clouds through ROS topics.
    \item We give a theoretic analysis of data propagation throughout the network and per-node storage constraints.
    \item We provide a channel quality study.
\end{itemize}

\section{Related Work}


Cooperative perception has advanced through simulation frameworks such as OPV2V \cite{xu2022opv2v} and infrastructure-based LTE/C-V2X studies \cite{khaleghian2024lte}. These works highlight the benefits of multi-view fusion and provide scalable environments for testing algorithms, though their simulation based nature limits realism. Our testbed complements this line of research by enabling real-time SDR-based LiDAR sharing across ROS-managed nodes, providing a physical platform for validating collaborative perception.

SDRs have been widely used to prototype systems such as IEEE 802.15.4 \cite{bloessl2013gnuradio}, LTE \cite{hematian2017sdr}, and WLAN \cite{mathworks_80211sdr}, ranging from open-source to closed-source. These efforts typically address PHY/MAC layer communication but don't address containerization for general node types and backend data processing (e.g. point-cloud fusion). We complement this body of research by adapting 802.11a/g reference designs \cite{mathworks_80211sdr} to 802.11p and pairing them with Dockerized ROS nodes for V2V perception sharing. Additionally, we leverage Matlab's opensource software and plug-and-play capabilities.

Decentralized vehicular networking has been explored in areas such as blockchain security \cite{jabbar2022blockchain,almatari2024blockchain} and peer-to-peer file sharing \cite{benet2014ipfs}. These studies emphasize scalability, trust and highlight the potential for secure vehicle-to-vehicle communication leveraging blockchain technology.

\section{System Design and Implementation}

\begin{figure}
    \centering
    \includegraphics[width=\columnwidth]{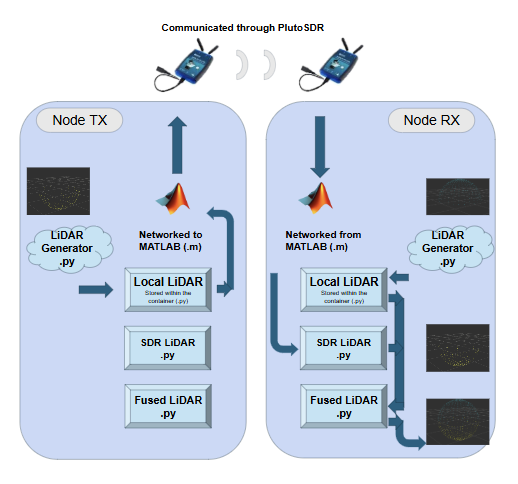}
    \caption{System overview of the IEEE 802.11p SDR V2V communication system. A single information pass shows how a node transmits and receives LiDAR point-cloud data and fuses it via ROS topics.}
    \label{fig:system}
\end{figure}

Our SDR-based V2V testbed is designed to be highly configurable and modular, supporting multiple modes of operation to accommodate different experimental setups and objectives. The system has four main modes:

\begin{itemize}[itemsep=1pt, topsep=2pt] 
    \item Transmit (TX) – Continuously sends generated LiDAR point cloud data over the air to other nodes.
    \item Receive (RX) – Listens for incoming transmissions from neighboring nodes and reconstructs the LiDAR data.
    \item Self-Transmit – Allows a single node to transmit and immediately receive its own signals for calibration or testing purposes.
    \item Discovery – Implements ALOHA-based medium access algorithm \cite{abramson1970aloha}, constantly listens but periodically broadcasts a presence signal at random intervals so that nodes can detect and identify each other in the network without message collisions.
\end{itemize}

These modes allow researchers to simulate real-world V2V scenarios, including one-to-one, one-to-many, and many-to-many interactions, without modifying the underlying hardware. The system is modular: both \texttt{Self-Transmit} and \texttt{Discovery} function as loops using the \texttt{TX} and \texttt{RX} modes as building blocks. The system integrates Matlab for SDR signal control and ROS containers for backend LiDAR data fusion. This architecture supports experimentation with distributed networks of vehicles, drones, or any node capable of running ROS in a Docker container and interfacing with SDR hardware.

\subsection{SDR Hardware Choice}
We use cost-effective ADALM-Pluto SDRs as the hardware backbone. SDRs provide full control over PHY and MAC layers, making them ideal for research and experimentation. These allow is to configure DSRC-compliant transmissions by adjusting parameters such as channel frequency, gain, and packet scheduling. This flexibility allows us to configure multiple network protocols through software (in our case Matlab) instead of having specialized automotive radios.

\subsection{IEEE 802.11p} 

The choice of IEEE 802.11p \cite{etsi302663-2013} or DSRC (Dedicated Short-Range Communication) is natural since it is the established standard for low-latency, short-range vehicular communications and most radios in "the wild" today aren't capable of communicating in more sophisticated CV2X or 5G formats. DSRC is specifically designed to support high-mobility environments and offers predictable latency and reliable packet delivery.

 IEEE 802.11p was introduced as an amendment to IEEE 802.11-2012 to enable communication \emph{outside the context of a basic service set (OCB)}. This eliminates time-consuming authentication and association, which are impractical in highly mobile vehicular environments. Instead, frames are transmitted with a wildcard BSSID, allowing true ad hoc V2V operation.

\paragraph{PHY Layer.}
The 802.11p PHY uses orthogonal frequency division multiplexing (OFDM) in the 5.9 GHz band with 10 MHz wide channels, halving the 20 MHz channelization of WLAN to improve Doppler tolerance. Each channel supports eight data rates from 3 Mbit/s to 27 Mbit/s, using BPSK, QPSK, 16-QAM, and 64-QAM with coding rates from 1/2 to 3/4 (see Table~B.1 in~\cite{etsi302663-2013}). Our implementation uses MCS 4, 16-QAM modulation with OFDM and code rate 3/4, providing 18 Mbit/s while maintaining resilience to multipath fading. OFDM enables robust multipath performance, critical for driver environments, while the choice of 16-QAM balances throughput and error resilience for LiDAR point-cloud transmissions.

\paragraph{MAC Layer.}
The 802.11p MAC is derived from IEEE 802.11e Enhanced Distributed Channel Access (EDCA), itself an extension of CSMA/CA with QoS support (refer to B.4.1 in \cite{etsi302663-2013}). In this scheme, a node performs carrier sensing and defers transmission by a random back-off drawn from a contention window. The window expands exponentially upon collisions and resets after successful transmissions. For broadcast V2V messages, acknowledgements (ACKs) are suppressed, so the backoff is only applied once per packet and the contention window remains at $CW_{min}$. This ensures low latency but provides no retransmission, a trade-off consistent with safety-critical broadcast beacons.  

\paragraph{Implementation note.}
While our PHY configuration aligns with 802.11p, our MAC protocol isn't the original 802.11p distributed CSMA/CA. Instead, our framework inherits form the Mathworks WLAN SDR design \cite{mathworks_80211sdr}, which is based on 802.11a/g. In the set up data is packaged into MAC service data units (MSDUs), converted into MAC protocol data units (MPDUs) and passed to the PHY layer as PSDU, similar to WLAN but not implementing full contention-based medium access. Hence, the system emulates time-sequence transmission rather than decentralized channel contention. The distinction is important to note, although the PHY layer matches DSRC/IEEE 802.11p, the MAC is a simplified WLAN-derived scheduler rather than full CSMA/CA.

\subsection{Integration with ROS and Docker} 

We set up a Robot Operating System (ROS) containerized in Docker handling perception and communication tasks. ROS functions as a middle-ware for processing received LiDAR data and transmitting local sensor (or in our case generated) data across the network. Data exchange is done through ROS topics, providing publisher-subscriber interfaces.

A ROS node publishes its own internal \texttt{/local\_lidar} topic, which in our case provides generated data (though in practice it would come from a LiDAR sensor). This data is published to the \texttt{local\_lidar} ROS topic, which MATLAB subscribes to before sending it through a connected ADALM-Pluto. Once the Pluto receives a signal from other nodes or itself, MATLAB decodes it and republishes it via ROS topics, this time through \texttt{/sdr\_lidar}, which the ROS node is subscribed to. The \texttt{/fuse\_point-cloud} callback continuously listens to both \texttt{/local\_lidar} and \texttt{/sdr\_lidar} topics and applies time synchronization. When data from both sources arrives approximately simultaneously, the callback fuses them and publishes the combined point cloud to the \texttt{/fuse\_point-cloud} topic for visualization. In this setup, \emph{topics} serve as communication channels for exchanging messages, while \emph{callbacks} define the processing logic triggered upon message arrival.

The visualization of these topics is enabled through \texttt{rviz}, allowing real-time inspection of point-cloud fusion and message latency.

\section{Theoretical Framework}

In this section we abstract the vehicular network in order to analyze potential communication constraints during implementation.

\subsection{Vehicular Ad Hoc Network (VANET) Model}

Let $\mathcal{N} = \{ n_1, \dots, n_N \}$ be the set of $N$ vehicle nodes, each with an SDR and ROS.  
Each node $n_i$ communicates with neighbors $\mathcal{N}_i \subseteq \mathcal{N}$ within range $r$.  
The dynamic connectivity graph is
\begin{equation}
G = (\mathcal{N}, \mathcal{E}), \quad 
\mathcal{E} = \{ (n_i,n_j) \mid d(n_i,n_j) \le r \},
\end{equation}
where $d(n_i,n_j)$ is Euclidean distance.

\subsection{Peer-to-Peer Communication}

Let $x_i(t) \in \mathbb{R}^d$ be node $i$'s state vector (e.g., LiDAR data).  
From eq. 7 in \cite{olfati2007consensus} a consensus update is
\begin{equation}
\small
x_i(t{+}1) = x_i(t) + \sum_{j \in \mathcal{N}_i} w_{ij}\big(x_j(t)-x_i(t)\big),
\end{equation}
with $w_{ij} \ge 0$, $\sum_{j\in\mathcal{N}_i} w_{ij}\le 1$.

As noted in \cite[Sec. III, Lemma 5]{olfati2007consensus}, such updates guarantee that all node states remain within the convex hull of the initial states, which is useful when aggregating sensor data across multiple nodes.

\subsection{Data Convergence}

If $G(t)$ stays connected, $x_i(t)\!\to\! x^*$ as $t \to \infty$, where $x^*$ is a weighted average of all nodes’ initial states (interpretable as a shared LiDAR environment estimate).  
The convergence rate depends on the algebraic connectivity $\lambda_2$ of Laplacian $L=D-A$ (eq. 9 in \cite{olfati2007consensus}).  

\subsection{Decentralized Storage (IPFS/Filecoin)}


IPFS \cite{ipfs_docs, benet2014ipfs} stores data as a content-addressed Merkle-DAG (UnixFS); files are chunked (typically $\approx$256\,KiB), each chunk is addressed by its cryptographic hash (a CID), discovered via a distributed hash table (DHT), and exchanged with peers using a block protocol (Bitswap). Filecoin adds persistent, incentivized storage with verifiable proofs (PoRep/PoSt) and replication via storage deals~\cite{filecoin2020}.

Let the data size of one LiDAR frame $V_{\text{frame}}\!\in\!\mathbb{R}^+$ be split into $M\!\in\!\mathbb{N}$ chunks with average size $v_{\text{chunk}}\!\in\!\mathbb{R}^+$ so that $V_{\text{frame}}\!\approx\!M v_{\text{chunk}}$. Node $i$ retains a local fraction $\alpha_i\!\in\![0,1]$ and has max storage capacity $S_i^{\max}$:

\begin{equation}
S_i \;=\; \alpha_i V_{\text{frame}} \;\le\; S_i^{\max}\in\mathbb{R}^+.
\end{equation}

Missing chunks are fetched on demand from a peer set $\mathcal{R}_i$ of VANET or IPFS nodes. Let $C_{ij}\!\in\!\mathbb{N}$ be the number of chunks of the current object cached from peer $j\!\in\!\mathcal{R}_i$. The instantaneous effective storage  $S_i^{\text{eff}}$ is

\begin{equation}
S_i^{\max} \;\le\; S_i^{\text{eff}} \;=\; S_i \;+\; v_{\text{chunk}} \sum_{j\in\mathcal{R}_i} C_{ij}.
\end{equation}

\subsection{Latency and Propagation}
Let $\bar{k}$ be the average node degree:
\begin{equation}
\bar{k} = \frac{2|\mathcal{E}|}{N} \approx \pi r^2 \rho,
\end{equation}
where $\rho$ is node density and $r$ is communication range (eq. 16 ~\cite{bettstetter2002connectivity}).

Let $\tau_{ij}$ be per-link latency. The network propagation delay for data across $N$ nodes with average degree $\bar{k}$ is
\begin{equation}
T_\text{prop} \approx \frac{N}{\bar{k}} \mathbb{E}[\tau_{ij}].
\end{equation}

Fetching remote chunks adds storage-aware latency, modeled as
\begin{equation}
T_{\text{fetch}} = T_{\text{lookup}} + \frac{v_{\text{chunk}}}{B_{\text{net}}} + T_{\text{decode}},
\end{equation}
where $T_{\text{lookup}}$ is DHT/provider discovery, $B_{\text{net}}$ is effective network throughput, and $T_{\text{decode}}$ accounts for decompression.

Together, $T_\text{prop}$ and $T_\text{fetch}$ quantify the end-to-end delay for distributed LiDAR sharing in the VANET and IPFS.

\subsection{Scalability and Trade-offs}

Higher $\bar{k}$ improves connectivity and consensus robustness but increases per-node communication. Storage scales with $N$ and $\bar{k}$, so nodes must balance redundancy and memory by dropping/compressing old data, limiting storage to local neighborhoods, or offloading to distributed systems like IPFS/Filecoin~\cite{benet2014ipfs,filecoin2020}.

\subsection{Global Vision}

Nodes may include vehicles, robots, drones, and infrastructure.  
Local VANETs perform consensus, share summaries, and federate into a global, decentralized perception network.

\section{Experiments and Evaluation}

\subsection{Setup}

To reproduce the experiments in practice, detailed setup instructions are provided in the repository's \texttt{README.md}.

\subsection{LiDAR Data Fusion}  

The following example demonstrates how the Docker container running ROS fuses and displays LiDAR data.

\begin{figure}[H]
    \centering
    \includegraphics[width=1\linewidth]{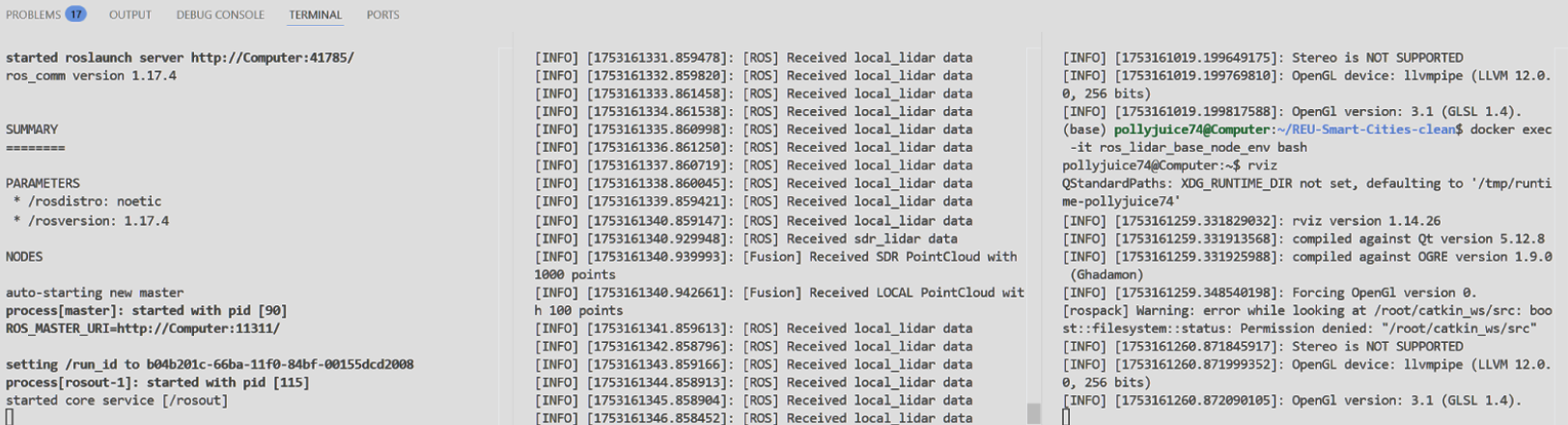}
    \caption{Three panels inside the same Docker container running ROS, the left panel initializes the roscore node, the middle panel runs and logs topic streams from our \texttt{LidarFusionNode} python script, the right panel runs a rviz instance visualizing received topic stream data.}
    \label{fig:placeholder}
\end{figure}

\begin{figure}[H]
    \centering
    \includegraphics[width=0.75\linewidth]{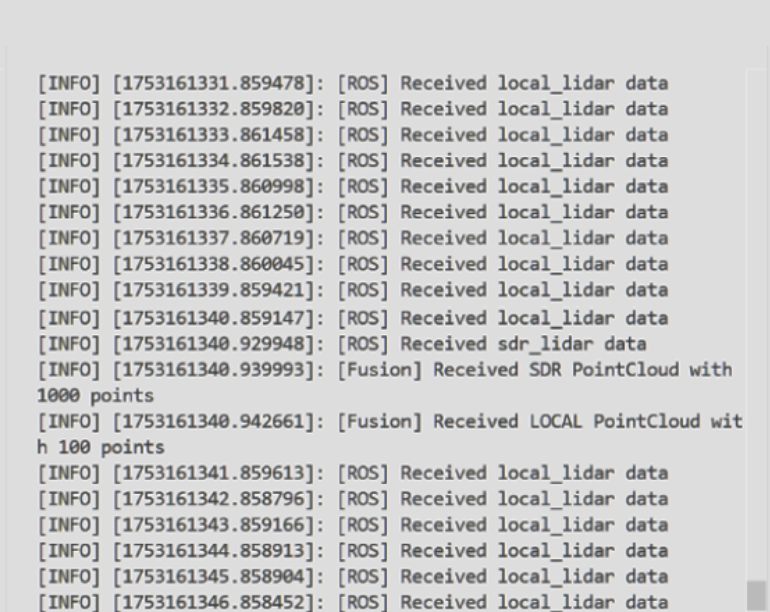}
    \caption{Close up of the middle panel showing a successful point-cloud fusion log.}
    \label{fig:placeholder}
\end{figure}

\begin{figure}[H]
    \centering
    \includegraphics[width=1\linewidth]{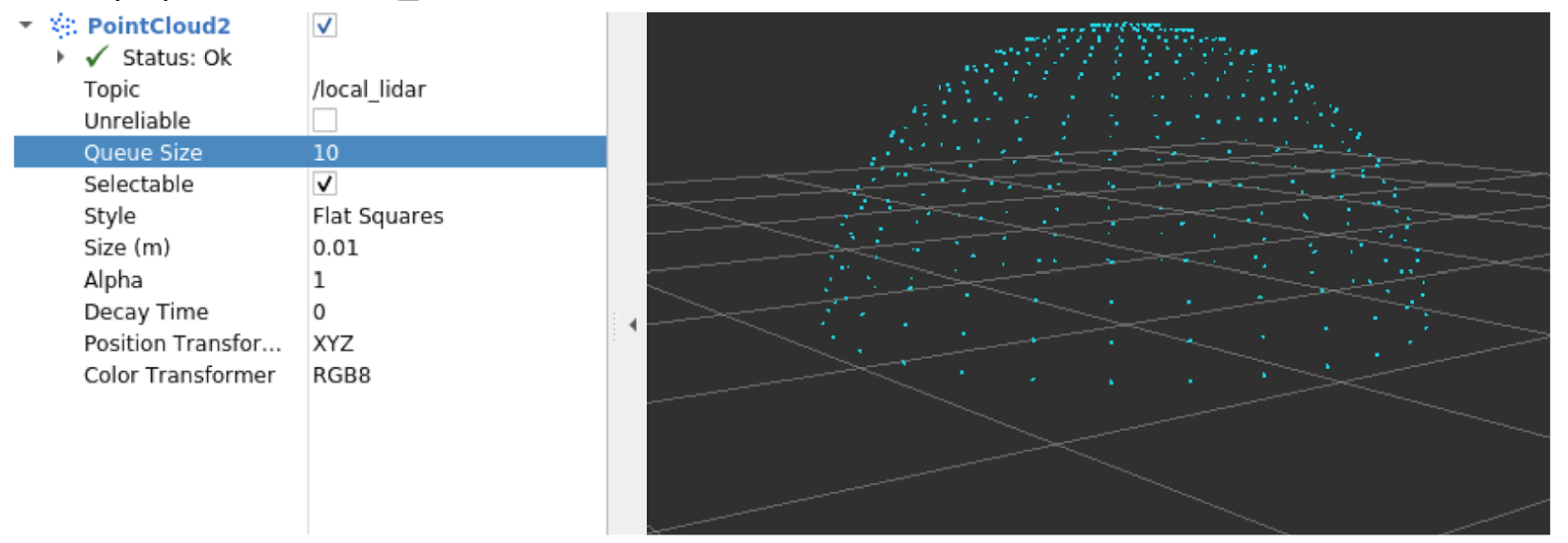}
    \caption{Point cloud topic stream visualization for \texttt{/local\_lidar} on rviz. }
    \label{fig:placeholder}
\end{figure}

\begin{figure}[H]
    \centering
    \includegraphics[width=1\linewidth]{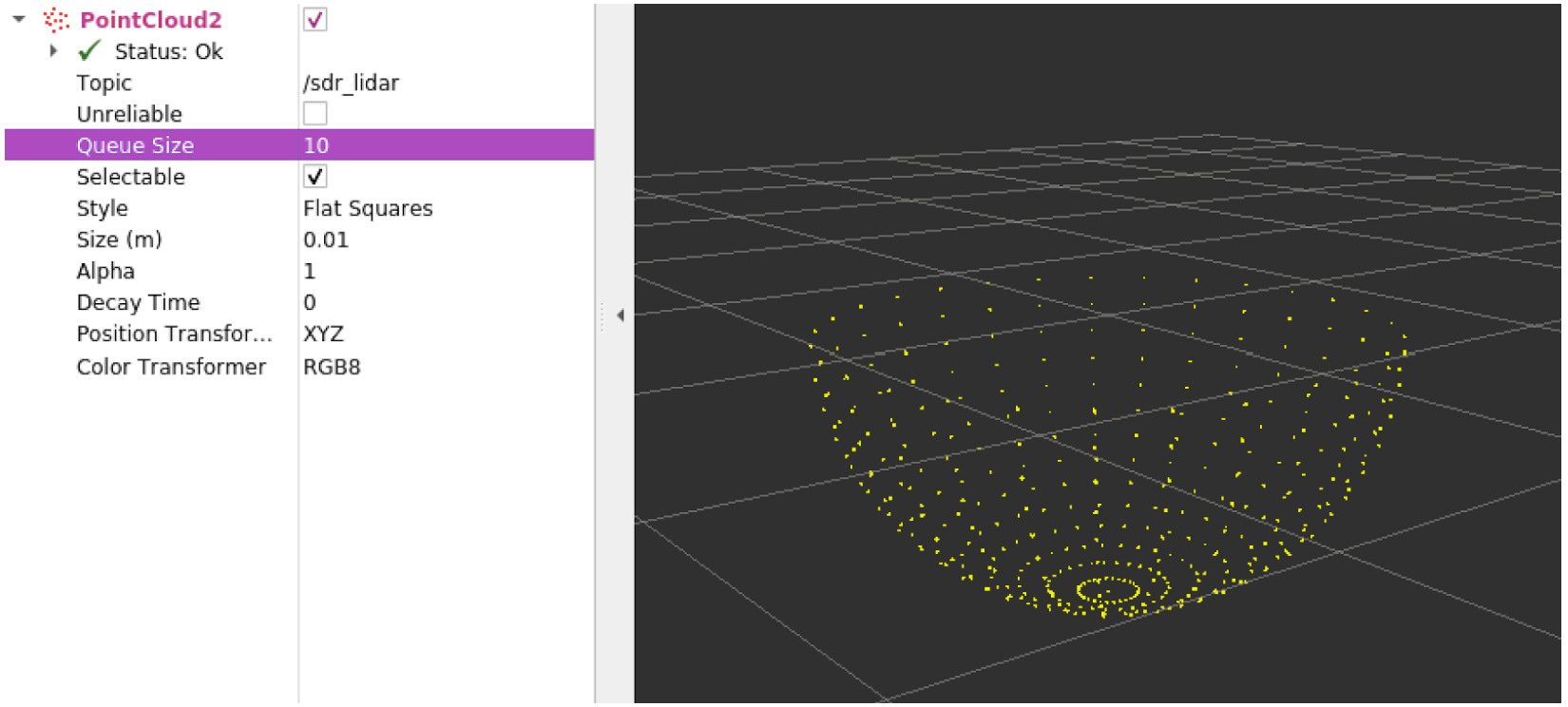}
    \caption{Point cloud topic stream visualization for \texttt{/sdr\_lidar} on rviz.}
    \label{fig:placeholder}
\end{figure}

\begin{figure}[H]
    \centering
    \includegraphics[width=1\linewidth]{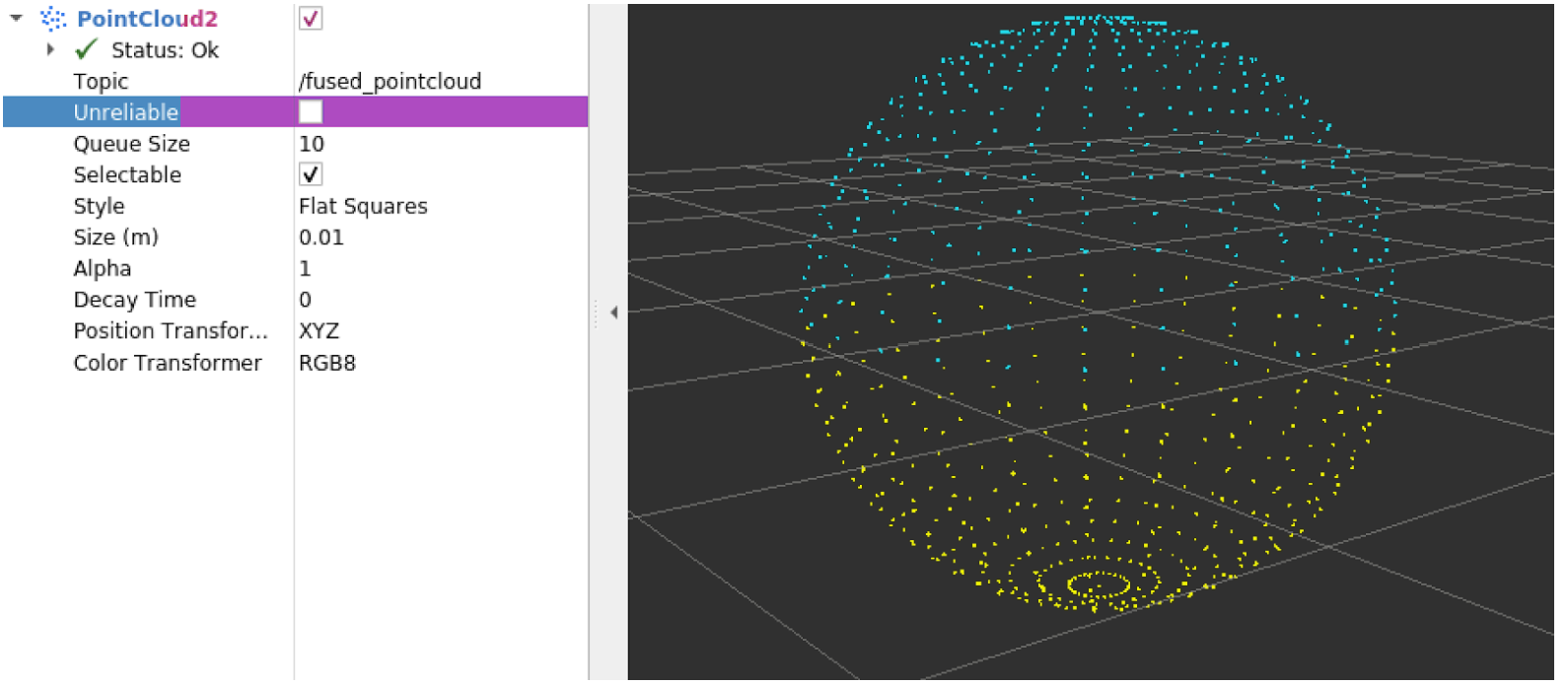}
    \caption{Point cloud topic stream visualization for \texttt{/fused\_point-cloud} on rviz.}
    \label{fig:placeholder}
\end{figure}

\subsection{Channel Quality Evaluation}

For channel quality evaluation we used three main metrics, a constellation diagram, a base-band signal spectrum and a bit error rate (BER) timeseries plot. The constellation diagram utilizes 16-QAM symbols plots, where each transmitted symbol should land on a discrete ideal point in the constellation for good channel quality. The base-band signal spectrum shows the frequency-domain representation of the received signal after down-conversion. The "box-like" lobe spans the DSRC bandwidth (10 MHz) and it's sharp edges indicate proper filtering, the dip in the center corresponds to the unused DC subcarrier in OFDM. The plot in kHz makes visualization easier. The bit error rate plot directly quantifies decoding accuracy over time (the code we used had convolutional decoding pre-built to estimate bit errors from a received signal).



Our first experiment evaluated the BER of the TX/RX modes. We evaluated four possible combinations of a Pluto transmitting to itself and to another Pluto over wire and antennas sending point-cloud data. When the Pluto communicated to itself it ran the RX mode on a process and then used the TX mode to transmit a single message all in one call. When the Pluto communicated to another Pluto, the receiveing SDR was in a RX loop while the transmitting SDR only called the TX once (1 foot apart). Each experiment was run ten times after warmup calls. We used manual gain control with over the wire values of 10, -10 and antenna values of 30, 0 for RX, TX respectively. It is worth mentioning that the inherited code had some convolutional error correction.

\begin{table}[h!]
\centering
\caption{Mean Bit Error Rate (BER) for different TX/RX modes with ADALM-Pluto SDRs.}
\label{tab:channel_quality}
\begin{tabular}{|p{1.8cm}|p{3.2cm}|p{1.6cm}|}
\hline
\textbf{Mode} & \textbf{Setup} & \textbf{Mean BER} \\
\hline
Self TX/RX & Single Pluto, loop-back (wired) & $0$ \\
\hline
Self TX/RX & Single Pluto, over antennas & $<10^{-3}$ \\
\hline
Pluto-to-Pluto & Two devices, wired & $0$ \\
\hline
Pluto-to-Pluto & Two devices, over antennas & $<10^{-2}$ \\
\hline
\end{tabular}
\end{table}

The following figure evaluates BER of the Self-Transmit mode over time. Automatic Gain Control (AGC) is configured. We attribute the higher BER to timing synchronization issues, which similarly affect Discovery mode, since both must alternate in a loop between TX and RX transmission. Implementation time limitations (9 weeks) restricted further optimization of the timing control.

\begin{figure}[H]
    \centering
    \includegraphics[width=1.0\linewidth]{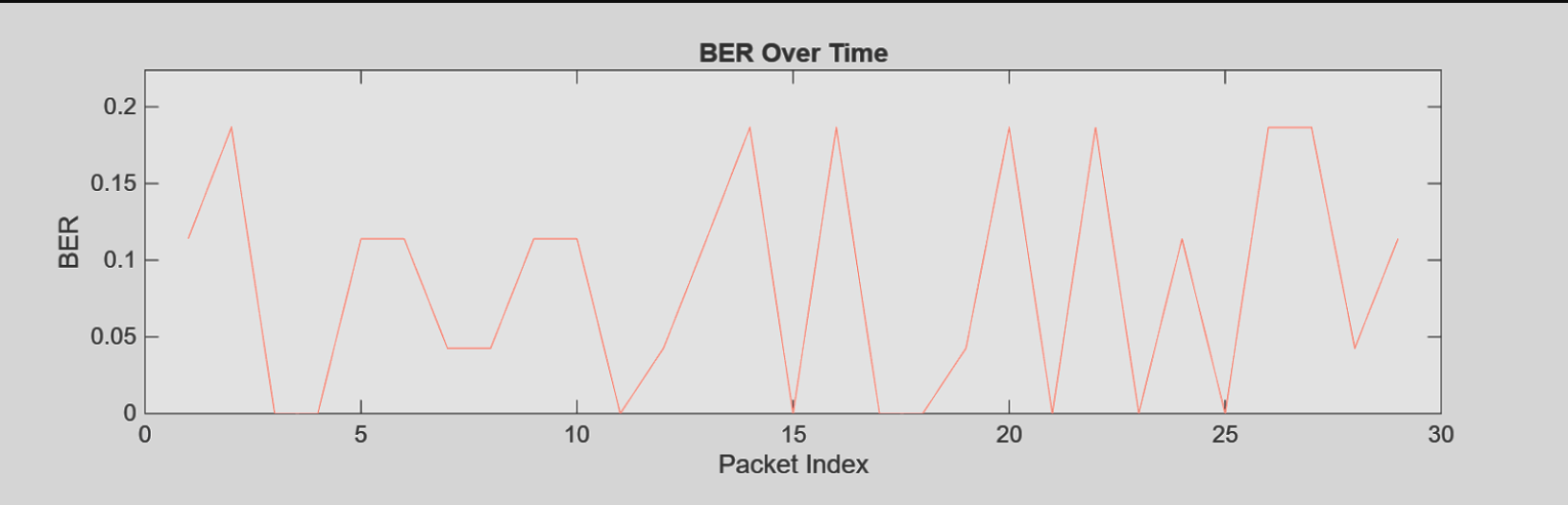}
        \caption{Self transmission over wire: 
        \newline
        Mean BER: 0.08381, Std: 0.07390
        \newline
        Mean Bits Received: 55,968, Packages Received: 29
        \newline
        Total Bits Received: 1,623,072
    }
    \label{fig:placeholder}
\end{figure}

The following plots are from a single communication link between two computers running our software. For the sake of demonstration, this example covers both Self-Transmit and Discovery modes. A video of the demo is available on our github repo.  

The Self-Transmit mode proves data is being transmitted and received (very useful for debugging). The Discovery mode proves that our system can transmit and receive data (like a half-duplex) and is not simply reading it's own transmitted data. The ADALM-Pluto antennas were roughly 4 feet apart (DSRC is for short range communication and it's range can be amplified using more production oriented hardware like USRPs). 


\begin{figure}[H]
    \centering
    \includegraphics[width=1.0\linewidth]{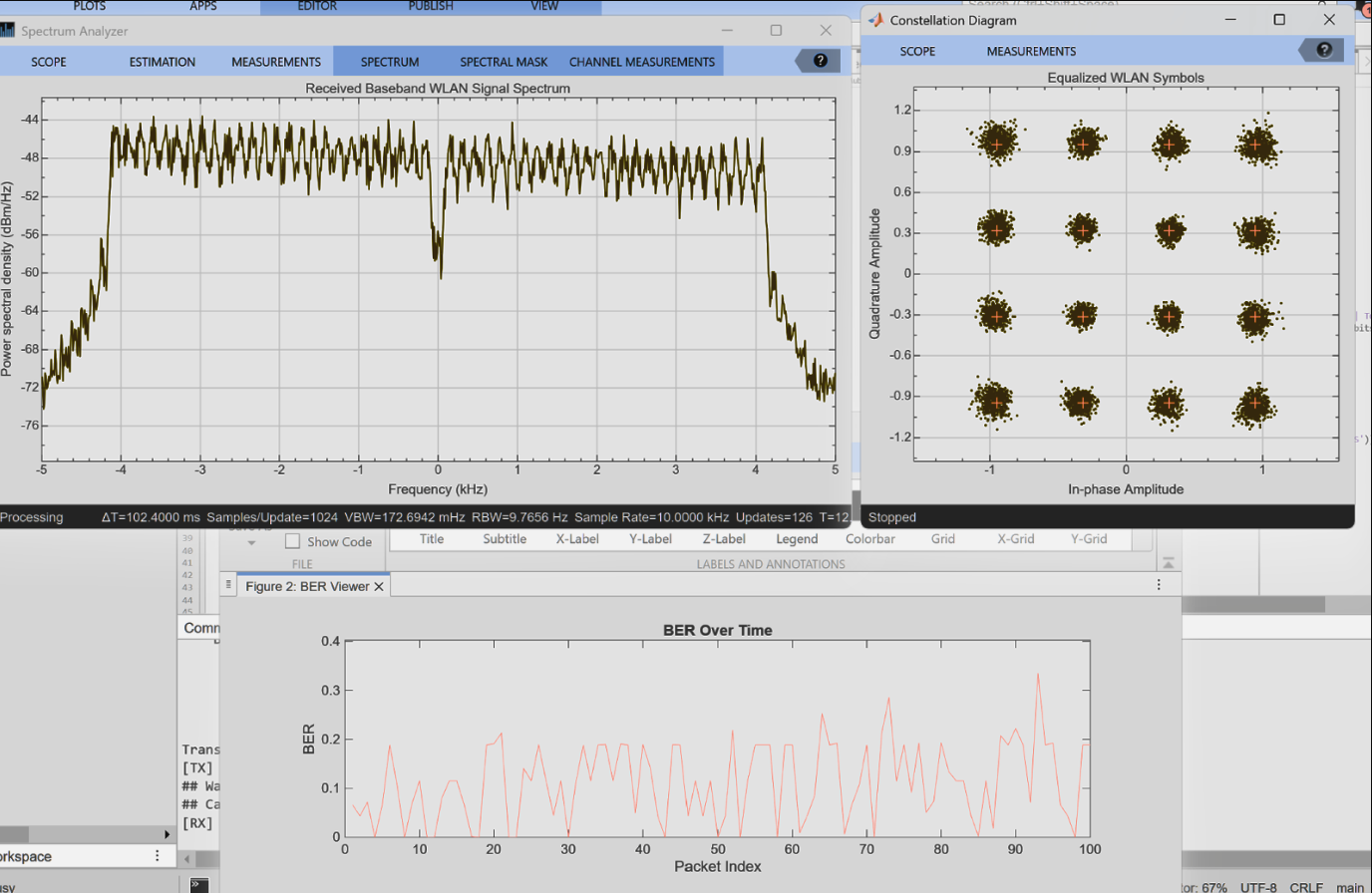}
    \caption{\parbox{\linewidth}{Self-transmit mode on computer 1:\\
        Mean BER: 0.11331, Std: 0.07967\\
        Mean Bits Received: 55,968, Packages Received: 100\\
        Total Bits Received: 5,596,800
    }}
    \label{fig:self_tx_antenas}
\end{figure}

\begin{figure}[H]
    \centering
    \includegraphics[width=1.0\linewidth]{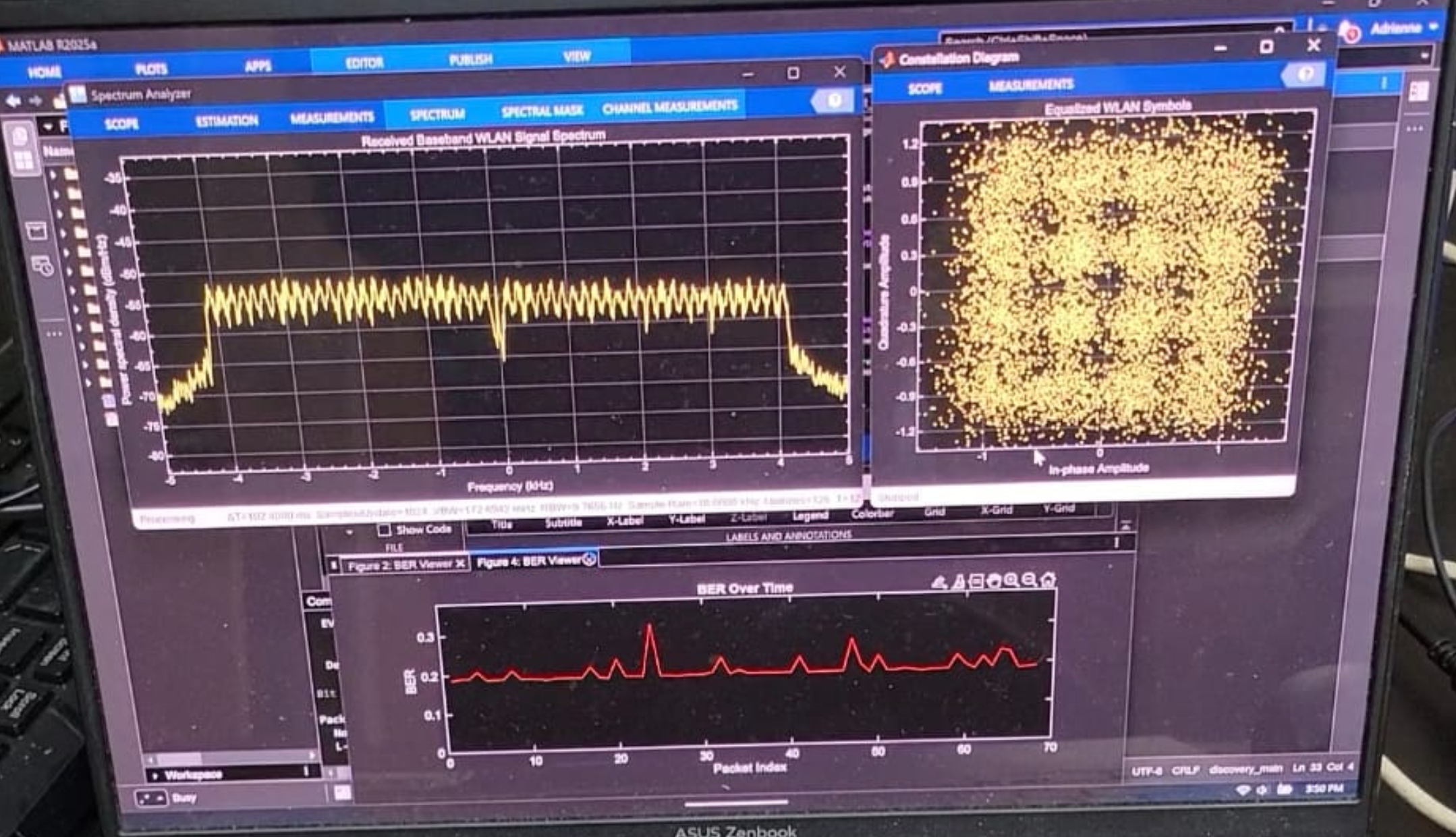}
    \caption{\parbox{\linewidth}{Discovery mode on computer 2:\\
        Mean BER: 0.18872, Std: 0.03181\\
        Mean Bits Received: 55,968, Packages Received: 65\\
        Total Bits Received: 3,637,920
    }}
    \label{fig:discovery_mode}
\end{figure}

\subsection*{Future Work}

A few avenues of research remain open for exploration:

\begin{itemize}[itemsep=1pt, topsep=2pt] 
    \item \textbf{Timing synchronization and networking layer:} In order to have proper node-to-node communication a robust networking layer is crucial.
    \item \textbf{Enhanced MAC protocols:} Future work could investigate custom MAC schemes that account for high LiDAR data volumes.
    \item \textbf{Edge-assisted and hybrid architectures:} Integrating Roadside Units (RSUs) or edge servers could support hierarchical consensus and reduce redundant transmissions.
    \item \textbf{Storage optimization:} Implementing distributed storage frameworks such as IPFS, along with LiDAR-specific compression techniques, are crucial for scalability.
    \item \textbf{Integration with cooperative perception algorithms:} Beyond data exchange, the testbed can serve as a foundation for implementing fusion strategies (early, late, or intermediate) and their impact on safety and autonomy.
\end{itemize}

\section{Conclusion}

We presented a modular, SDR-based IEEE 802.11p testbed for distributed Vehicle-to-Vehicle communication, enabling real-time LiDAR point cloud sharing and fusion across ROS-managed nodes. Through the combination of Matlab-based SDR control and Dockerized ROS nodes, we bridge the gap between simulation and real-world experimentation. The testbed supports heterogeneous device-node creation, channel quality metrics and demonstrates scalable perception sharing. We provide a theoretical framework for how this network's constraints could be analyzed.

\section*{Acknowledgments}

The authors would like to thank NSF Grant No. 2349616 for providing funds and the UNLV Smart Cities REU for hosting the research internship. We extend our gratitude to Dr. Peter Stubberud, and Dr. Ebrahim Saberinia for their dedicated mentorship and guidance throughout the project. We also thank Dr. Brendan Morris for leading the REU as well as providing valued feedback on the research. 



\printbibliography

\end{document}